# Properties of recent IBAD-MOCVD Coated Conductors relevant to their high field, low temperature magnet use

**V Braccini[1*], A Xu[1], J Jaroszynski[1], Y Xin[1], D C Larbalestier[1], Y Chen[2], G Carota[2], J Dackow[2], I Kesgin[3], Y Yao[3], A Guevara[3], T Shi[3] and V Selvamanickam[3]**

[1] Applied Superconductivity Center, National High Magnetic Field Laboratory, 2031 E. Paul Dirac Dr., Tallahassee, FL 32310
[2] SuperPower Inc., 450 Duane Ave., Schenectady, NY 12304
[3] Department of Mechanical Engineering and the Texas Center for Superconductivity at the University of Houston, 4800 Calhoun Rd, Houston, TX 77204

E-mail: braccini@asc.magnet.fsu.edu

**Abstract** $BaZrO_3$ (BZO) nanorods are now incorporated into production IBAD-MOCVD coated conductors. Here we compare several examples of both BZO-free and BZO-containing coated conductors using critical current ($I_c$) characterizations at 4.2 K over their full angular range up to fields of 31 T. We find that BZO nanorods do not produce any *c*-axis distortion of the critical current density $J_c(\theta)$ curve at 4.2 K at any field, but also that pinning is nevertheless strongly enhanced compared to the non-BZO conductors. We also find that the tendency of the *ab*-plane $J_c(\theta)$ peak to become cusp-like is moderated by BZO and we define a new figure of merit that may be helpful for magnet design – the OADI (Off-Axis Double $I_c$), which clearly shows that BZO broadens the *ab*-plane peak and thus raises $J_c$ 5-30° away from the tape plane, where the most critical approach to $I_c$ occurs in many coil designs. We describe some experimental procedures that may make critical current $I_c$ tests of these very high current tapes more tractable at 4.2 K, where $I_c$ exceeds 1000 A even for 4 mm wide tape with only 1 μm thickness of superconductor. A positive conclusion is that BZO is very beneficial for the $J_c$ characteristics at 4.2 K, just as it is at higher temperatures, where the correlated *c*-axis pinning effects of the nanorods are much more obvious.

## 1. Introduction

It is now clear that High-Temperature Superconductors (HTS), especially $YBa_2Cu_3O_x$ or more general $REBa_2Cu_3O_x$ (REBCO, where RE = rare earth element) coated conductors (CCs), will serve a growing demand for very high field magnets. Power applications of superconductors such as motors, generators and transmission lines require wires exhibiting high pinning forces at high temperatures [1]: a huge amount of work has been successfully performed in the last years with the aim of improving the effective pinning at liquid nitrogen temperature through engineering of defects in REBCO matrix [2 and ref. therein]. The construction of magnets for generation of fields above 30 T though requires cooling down to 4.2 K: the flux pinning mechanisms may be very different at high and low temperatures – i.e. close to the critical temperature (> $T_c/2$) or well below it (< $T_c/2$) - and, while CCs are widely characterized at 77 K or in general at relatively high temperatures, there is only a small literature at liquid He temperatures. At 4.2 K the vortex core size is smaller and thermal fluctuation effects are strongly reduced, allowing weaker pins to become more important. Furthermore, critical current measurements are much more complicated at low temperatures, because much higher currents and magnetic fields are required.

Dopants which have been shown to be very successful in improving the critical current behaviour in field $I_c(H)$, as well as the pinning performances of REBCO conductors at high temperatures, are $BaZrO_3$ (BZO) nanoparticles. BZO was first incorporated in films grown by Pulsed Laser Deposition [3], leading to a substantial enhancement of the critical current at all the orientations with respect to the magnetic field and in particular to a stronger pinning along the *c*-axis. A chemical solution deposition technique was applied to generate a dense array of BZO nanodots randomly oriented and distributed within the YBCO matrix, leading to a quasi-isotropic strong vortex-pinning landscape [4]. Later on, the scalable and economically viable technique of metal-organic chemical vapour deposition (MOCVD) has been successfully employed to fabricate REBCO tapes with enhanced pinning performance through the formation of columnar stacks of BZO nanodots [5-7].

[*] Permanent address: CNR-SPIN, C.so Perrone 24, I-16152 Genova, Italy

However, the measurements described in the literature were almost all performed at relatively high temperatures, about or slightly below liquid nitrogen temperature. Even so, while it is well established that REBCO can incorporate many different types of pinning centres with significant effect on their critical current density $J_c$, the angular anisotropy of the critical current density $J_c(\theta)$ and the microscopic mechanisms controlling $J_c$ are still poorly understood in these emerging conductors. A detailed knowledge of the angular dependence of $I_c(\theta,H)$ is crucial for magnet construction, since it is required to predict the coil quench point. Due to the strong anisotropy of $J_c$, the most critical part of the magnet is generally at the end of the coil where the radial field is high, in contrast to a conventional superconducting magnet made of isotropic superconductors, where the highest field point is also closest to $I_c$ and is generally found at the centre of the magnet.

The possibility of studying the angular and magnetic field dependence of the critical current up to very high fields with particular attention to the behaviour at 4.2 K is therefore important, both to understand the pinning mechanisms and to construct superconducting magnets, as well as for guiding the manufacturing of conductors with enhanced properties.

We present here detailed angular measurements of the critical current $I_c$ in different CCs fabricated by SuperPower Inc., mainly focusing on the effect of BZO nanorods by a comparison between the superconducting properties of undoped CCs and CCs with BZO precipitates. The measurements were performed in magnetic fields up to 31 T at 4.2 K so as to be relevant for very high field magnet design.

We find that strongly correlated pins like BZO substantially modify the $I_c(\theta,H)$ angular dependencies at 4.2 K. In particular, BZO doping markedly widens the cusp-like $I_c$ maxima around the CC plane at all magnetic fields. Thus BZO-doped CCs are more suitable for magnet construction than undoped CCs, despite their faster decrease of $I_c$ with increasing magnetic field in the off-plane configuration.

## 2. Experimental details

The CCs characterized here are REBCO HTS wires produced by SuperPower Inc. They consist of biaxially-textured buffers made by ion beam assisted deposition (IBAD) of MgO on high-strength Hastelloy substrates followed by REBCO (RE = Y, Sm, Gd or a mixture of them) film metal organic chemical vapour deposition, as described in more detail elsewhere [8-10]. The most recent CCs were prepared by adding Zr-tetra methyl heptanedionate (thd) to the precursor solution for Zr doping [6]: BZO nanocolumns formed by a self-assembly process were found to be responsible for improved pinning [3]. Zr content was optimized at a level of 7.5 mol % in the newer samples. Earlier tapes made without Zr doping contained Y and Sm as the rare earth constituent. Recent tapes made without Zr doping contained only Gd as the rare earth constituent. The composition of all Zr-doped tapes except one consisted of Y and Gd rare earths in equal mole fractions. The nominal thicknesses of the CCs were mostly in the vicinity of 1 μm, although two were about 2 μm and one was much thinner. In table 1 the main parameters of the 10 samples studied in this paper are reported.

**Table 1** Main parameters of the CCs.

| *Sample designation* | | *Dopant* | *Mol%Zr* | *RE composition* | *Electroplated Cu thickness* | *RE123 thickness (μm)[a]* |
|---|---|---|---|---|---|---|
| M3-513 FS | SP02 | Non BZO | - | $Y_{1.3}Sm_{0.2}$ | 20μm/20μm Cu | 2.1 |
| M3-609 MS | SP06 | Non BZO | - | $Gd_{1.5}$ | 20μm/20μm Cu | 2.1 |
| M3-522 MS | SP07 | Non BZO | - | $Y_{1.3}Sm_{0.2}$ | 20μm/20μm Cu | |
| M3-646-2 FS | SP19 | Non BZO | - | $Gd_{1.5}$ | 20μm/20μm Cu | 0.85* |
| M3-612-19 BS | SP20 | Non BZO | - | $Gd_{1.5}$ | 20μm/20μm Cu | 1.05* |
| 2140 RD | | 'Old' BZO | 5% | $Gd_{0.65}Y_{0.65}$ | - | 0.35* |
| M3-687-2 MS | | 'Old' BZO | 10% | $Gd_{0.65}Y_{0.65}$ | 20μm/20μm Cu | 0.94 |
| 2265-11 RD | | 'New' BZO | 7.5% | $Gd_{0.6}Y_{0.6}$ | - | 0.9* |
| 2291-8 RD | | 'New' BZO | 7.5% | $Y_{1.2}$ | - | 0.9* |
| M3-745-20 MS | SP26 | 'New' BZO | 7.5% | $Gd_{0.6}Y_{0.6}$ | 50μm/50μm Cu | 1.1 |

[a] The thickness values were determined by electron microscopy examination, except in the samples marked with *, where the value expected from the deposition rate is taken

Four-probe, transport $I_c$ measurements were performed at 4.2 K with a home-built, 500 A rotator probe [11] in both a 52 mm cold bore, 15 T superconducting magnet and the NHMFL 31 T, 52 mm warm bore, Bitter magnet, fitted with a 38 mm bore He cryostat. The rotator sample platform is driven by a stepper motor pulling Kevlar strings strong enough to survive high torques and low temperatures. We define $\theta$ as the angle between the external magnetic field and the normal to the CC plane, thus $\theta = 90°$ is parallel to the tape plane. $\theta$ can vary between about -10° and +120° with respect to the magnetic field orientation. Its actual value is checked with a Hall probe glued below the sample. Current leads are made by 4 mm wide YBCO CC tape supported within Cu braids.

At 4.2 K, for the usual 4 mm wide CCs, $I_c$ easily exceeds 1000 A, especially when the field orientation is parallel to the *ab* plane. We therefore chemically etched away the surrounding Cu stabilizer of the conductor and the Ag coating when present, and made bridges of ~ 1 mm width. Actually we observed that most of the time, the critical current is about 10% higher when measured through a narrow bridge: we think that this might be due to the fact that the width of the conductor actually carrying the current is somehow less than 4 mm.

There are several problems to deal with while performing this type of measurement, some of which are also sample dependent. Some samples were weak and easily delaminated. Occasionally, *V-I* characteristics indicating significant current transfer resistance were seen. Resistive backgrounds strongly suggest a long current transfer length between the external Cu and the REBCO, sometimes causing considerable sample heating. In figure 1 we show two examples of 'bad' curves recorded on some of the CCs. In figure 1(a) we observe a significant resistive background. A high resistance somewhere between the Cu/Ag/REBCO and the current leads causes a lossy current transfer. The crucial parameter in this case is the Current Transfer Length (CTL), which depends on the resistivity of the Cu, Ag, REBCO and the interfaces through which current has to pass [12, 13]. When the interface resistance becomes high, the CTL is not negligible but the needed condition that the voltage tap separation is longer than the CTL cannot be fulfilled because we are limited by both the small 26 mm length of samples that we can accommodate in our rotator and the small area of the current contacts. After subtracting the linear background and any offset, it is generally possible to extract an $I_c$ value defined at the usual 1 μV/cm, although figure 1(b) shows an example of two *V-I* characteristics which are highly anomalous well below $I_c$. Anomalous voltage peaks have been observed below $I_c$ and are generally [14] ascribed to current redistributions provoked by sample inhomogeneity. Here we think a superposition of two issues occurs: 1. a bad current transfer which gives the initial linear behaviour and 2. overheating due to high resistance between current leads and superconducting phase. Indeed the resistive slope in figure 1(b) is significantly larger than in figure 1(a), leading to the overheating.

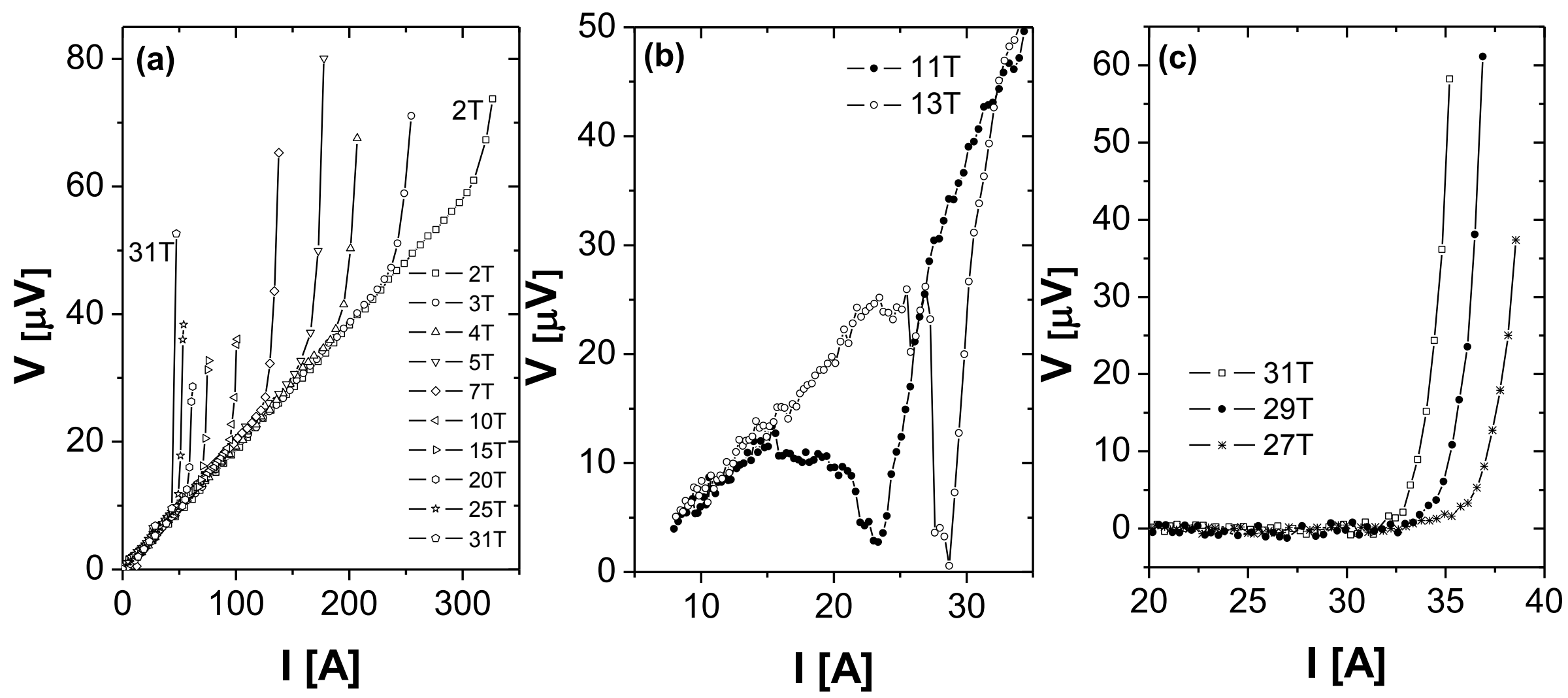

**Figure 1** Examples of 'bad' and 'good' *V-I* characteristics measured at 4.2 K at different magnetic fields on different samples. (a) Transitions with a resistive background due to bad current transfer along Cu/Ag/REBCO; (b) transitions with superimposed effect of high contact resistance and overheating of the sample; (c) 'good' *V-I* transitions.

From such transitions it is not possible to extract the actual $I_c$ value with good accuracy. As a comparison, in figure 1(c) 'good' transitions are shown, which do not exhibit any of the problems discussed above. Current-transfer voltages are not observed when samples coated only by silver are measured. This suggests that the copper-silver interface is often the source of these problems.

Usually samples were soldered on small printed circuit boards using indium, which makes it easier to support voltage contact wires, as well as to diminish their inductive loop. The latter is crucial to minimize the noise in strong magnetic fields. Even more important, indium-soldering apparently supports the samples against degradation, delamination and burnout, which plagued our early measurements. However, soldering (even with indium) often destroys samples made without electroplated Cu, necessitating multiple sample mounting trials. Sometimes very good *V-I* transitions were obtained just by pressing samples onto a thin G-10 slab or between two brass slices, either with REBCO in the face-up or face-down configuration.

Transmission Electron Microscopy (TEM) images were taken on a JEOL2011 microscope with a $LaB_6$ filament, and a point resolution of 2.3 Å at 200 kV. A cross-sectional TEM image of the sample SP26 is shown in figure 2(a), where the view along the *a* axis shows a high density of BZO nanorods distributed through the whole REBCO layer, while figure. 2(b) shows the nanorods viewed end on in a plan view TEM image along [001]. Figure 2(c) shows a high magnification view of several BZO nanorods. The average diameter of the nanorods is about 8 nm and the measured areal density is 8.4 x $10^8$ $mm^{-2}$, a density equivalent to a matching field $B_m = \phi_0/a_0^2 \sim 2.5$ T, where $a_0$ is the spacing between nanorods and $\phi_0$ the flux quantum (=2 $10^{-15}$ Wb). The nanorods have a splay angle of about 10 to 15 degrees around the *c* axis. Apart from the nanorods, other defects in the REBCO layer are $RE_2O_3$ precipitates that form in rows on the *ab* planes together with stacking faults. The $RE_2O_3$ precipitates have an average size of 9 nm diameter. Threading dislocations are also present and lie principally along the *c* direction.

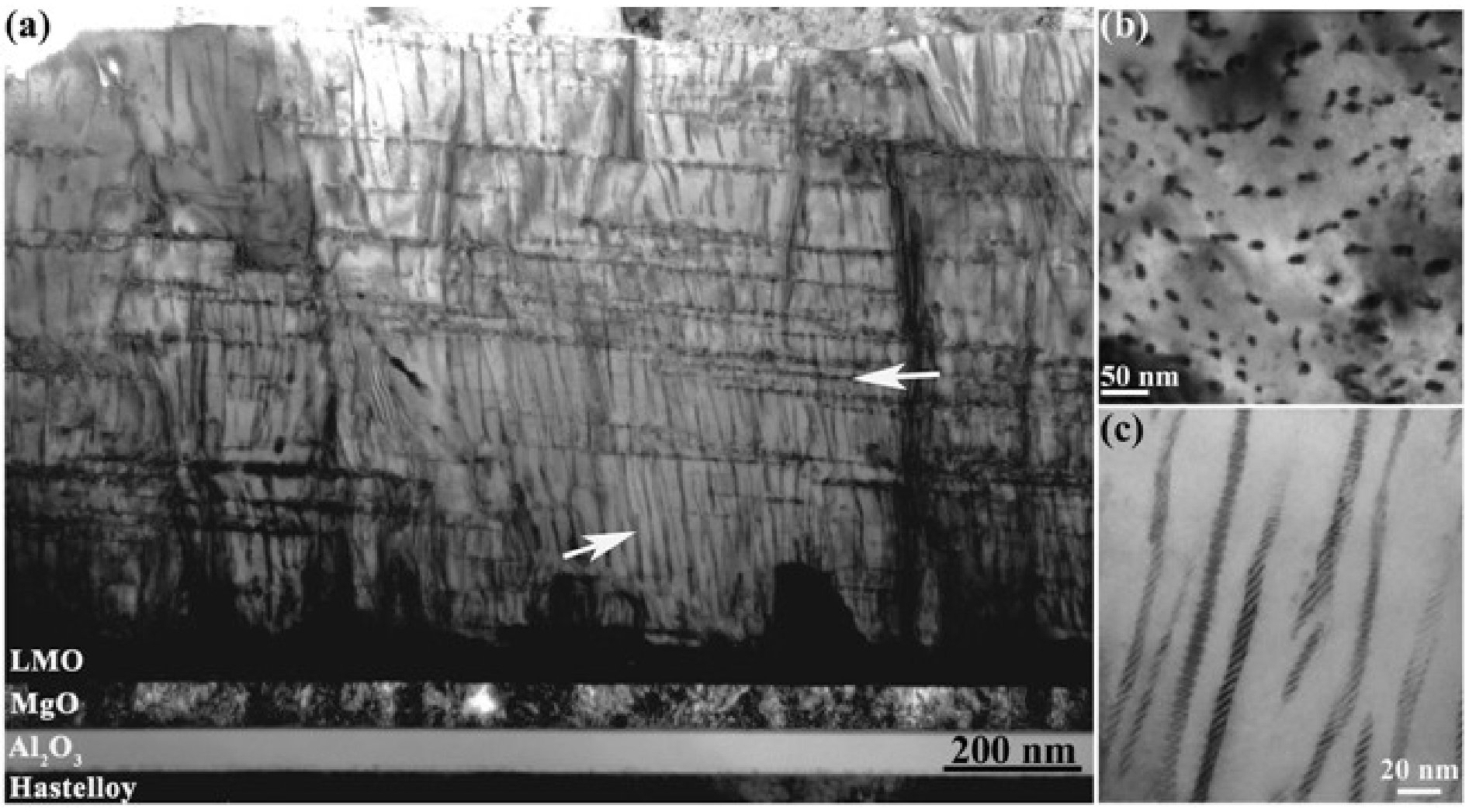

**Figure 2** TEM bright field images of the microstructure of sample SP26. (a) Cross sectional bright field image viewed along [100] showing the high density of nanorods through the whole REBCO layer (indicated by tilted arrow), small $RE_2O_3$ precipitates forming in line arrays on the *a-b* plane (indicated by horizontal arrow), and some threading dislocations. (b) Plan view bright field image viewed along [001] *c* axis showing the nanorods end on. (c) A magnified image of the BZO nanorods.

## 3. Results

CCs currently in production are optimized for use at 77 K where they are widely characterized, but their performance at high temperatures is not adequate for high field magnet application, for which cooling to 4.2 K is needed for > 30 T field generation. Manufacturers' efforts are mainly devoted to improving $I_c$ at 77 K, but this does not necessarily mean an improvement at 4.2 K, as is clearly shown in figure 3, where we report $I_c$ at 4.2 K, 14 T vs. $I_c$ at 77 K, self field for several samples. Roughly speaking, the samples lie on two different trend lines: the best samples at high temperatures have actually a lower $I_c$ at 4.2 K and vice versa. This has to be due to the different pinning or current-limiting mechanisms which come into play in the two regimes and different fields, as we will analyze in the following. We observe that some samples show a spread in the 4.2 K values, when different parts of the conductors are measured. In some cases this is related to an $I_c$ variation along the CC [15], while at other times it is due to a variation of $I_c$ across the width of the conductor which appears when comparing full 4 mm wide or 1-2 mm wide bridged samples, as previously outlined.

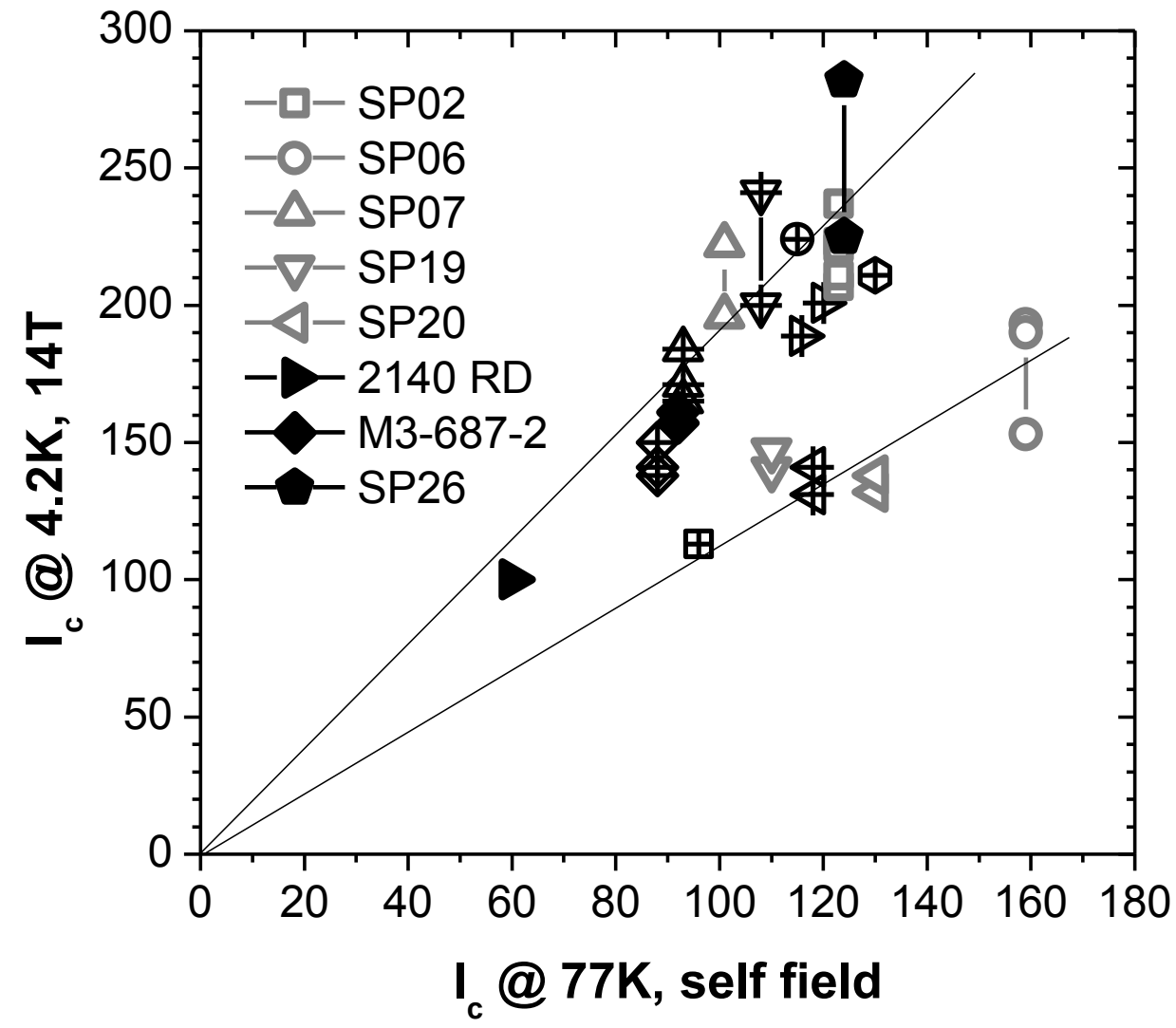

**Figure 3** $I_c$ at 4.2 K, 14 T vs. $I_c$ at 77 K, self field on several 4 mm-wide CCs. Together with some of the samples named in table 1 which are outlined in the legend, there are data on other CCs (hollow symbols with a cross). The spread in the 4.2 K data arises from the track width variability of $I_c$ in different samples, as mentioned in the text [15]. The two lines are guides for the eye. In this and future figures, hollow gray symbols indicate non-BZO samples, while full black symbols refer to BZO containing samples.

In figure 4 the critical current normalized to the standard 4 mm width of the CCs listed in table 1 is plotted for $H$ perpendicular to the $ab$ plane, as measured up to 31 T at 4.2 K. We can see that the $I_c$ values are quite dispersed and that the field dependence looks steeper for the samples with BZO additions.

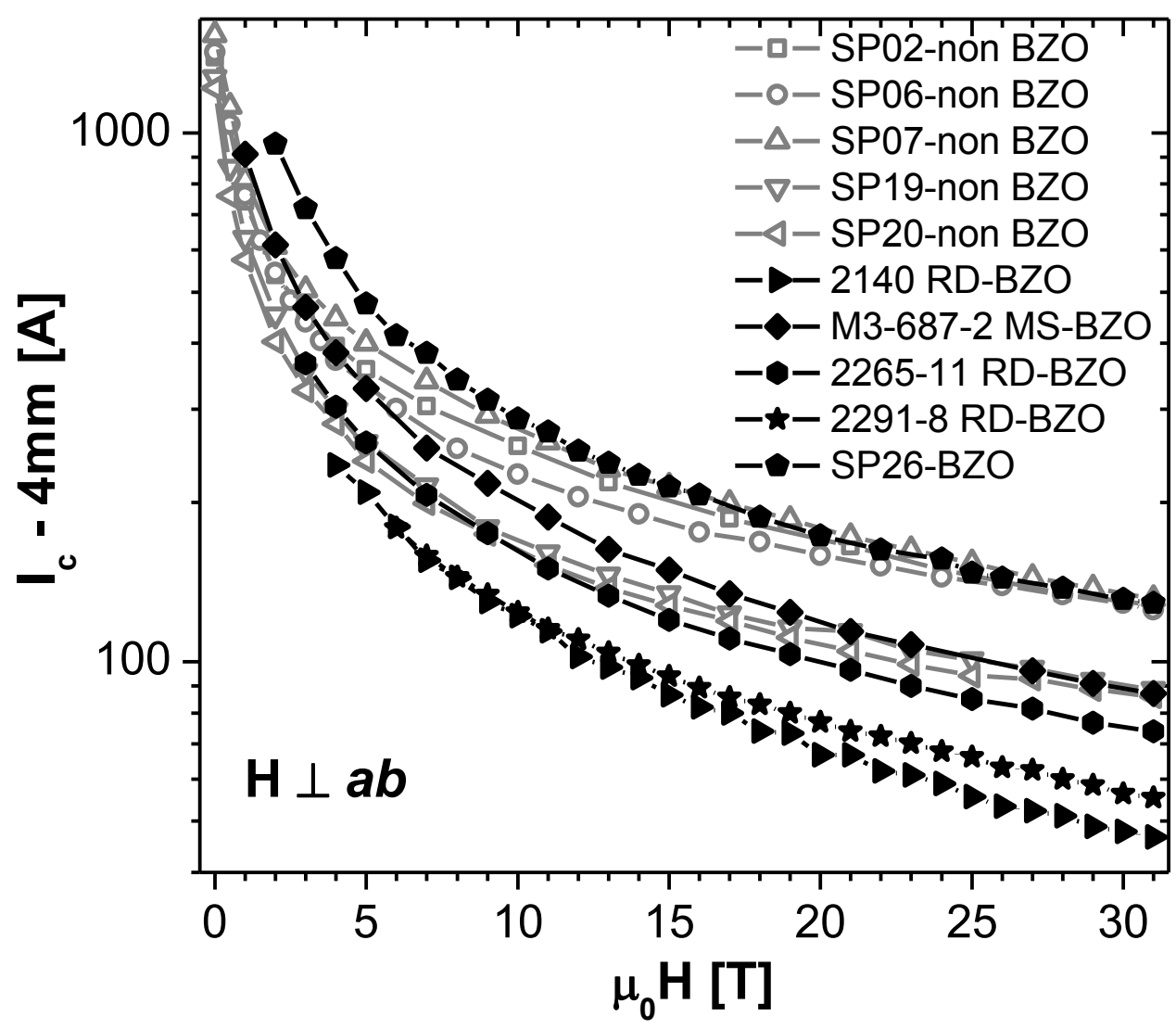

**Figure 4** $I_c$ vs. magnetic field at 4.2 K for $H$ perpendicular to the $ab$ plane.

In figure 5 $I_c(\theta,H)$ is reported for four of the samples listed in table 1, two - (a) and (b) - without and two - (c) and (d) - with BZO additions. In contrast to what is generally observed at 77 K [11], the behaviour of these samples is quite similar. This is at first sight rather surprising, because there is no obvious sign of the strongly correlated c-axis pinning seen at higher temperatures for the BZO samples, which strongly contrasts with the BZO-free conductors. In fact there is no *c*-axis maximum at any field, and a cusp-like *ab* peak is observed, which becomes more pronounced at high magnetic fields. If we look more carefully, we can see that BZO nanorods actually significantly widen the *ab* peak, leading to a higher current 5°-20° off the *ab* plane where it is valuable for magnet construction. In order to better visualize this widening of the *ab*-plane peak with BZO additions, we plot in figure 6 the angular dependence of the same four samples at 5, 10, 20 and 25 T, after normalizing $I_c$ to its value perpendicular to the *ab* plane.

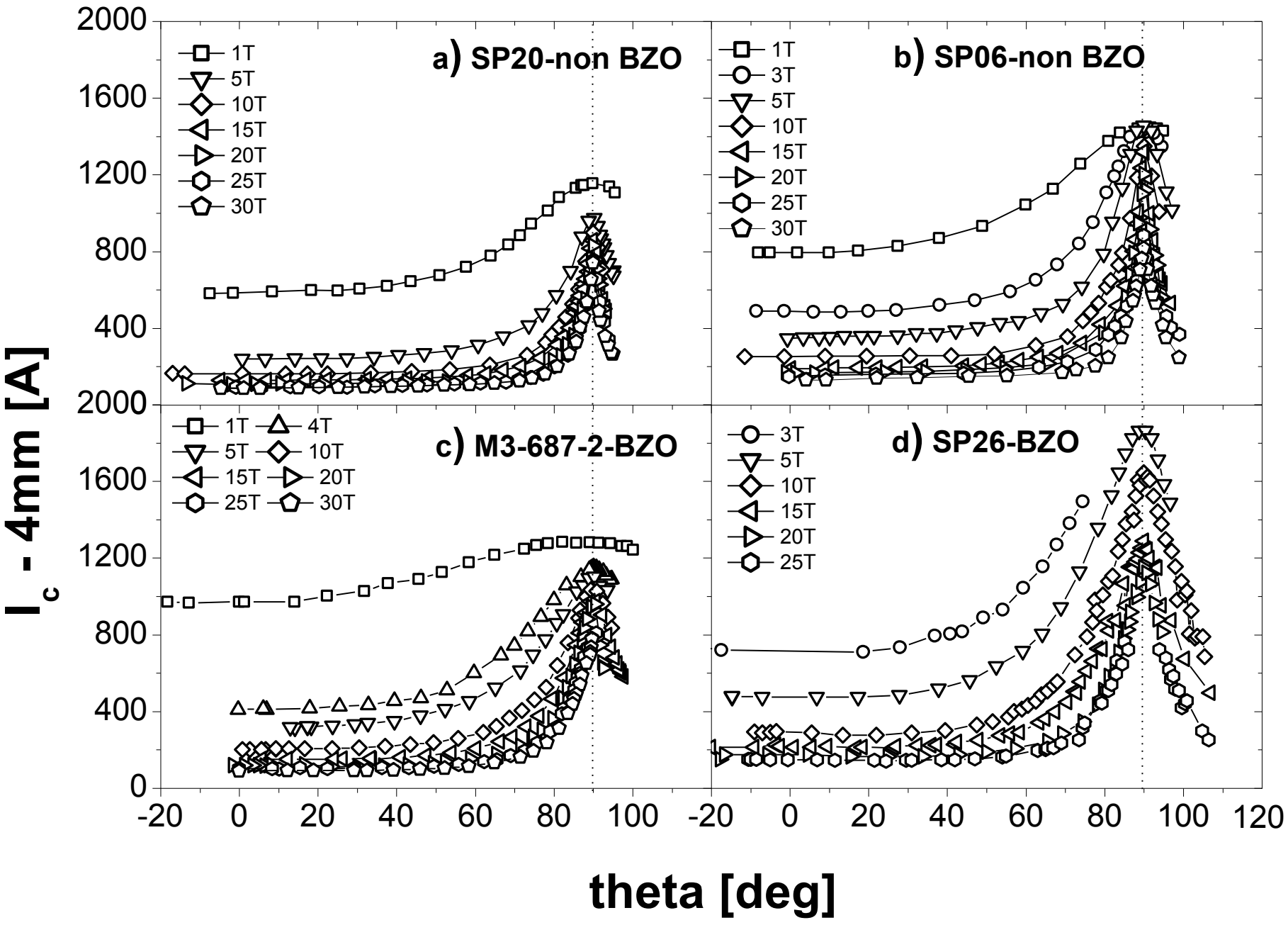

**Figure 5** $I_c$ angular dependence at 4.2 K and various magnetic fields up to 30 T for four of the samples listed in table 1: (a) and (b) without BZO nanorods, (c) and (d) with BZO precipitates. 90° is the configuration parallel to the CC plane.

The widening of the *ab*-plane peaks occurring when BZO nanorods are present, is shown in another way in figure 7, where we introduce a parameter defined as the Off-Axis Double $I_c$ (OADI) metric - i.e. the angle off the *ab* plane axis at which $I_c$ becomes twice the $I_c$ in the perpendicular orientation. This parameter – unlike the more commonly used FWHM about the *ab*-plane – is not affected by any unusually high (or low) *ab*-plane peak and it is thus more useful for magnet design. In fact, sometimes the maximum $I_c$ might be underestimated, due to heating effects or degradations which are likely to occur when the current is very high. A clear trend is observed in this plot: BZO nanoprecipitates substantially widen the *ab* peak, making the latest generation of BZO-added CCs much more appealing for magnet construction.

In figure 8 we compare the field dependence of $I_c$ of a well-performing non-BZO sample (SP06) with the new BZO CC (SP26) at three different field orientations: in plane, 5° off plane and 90° off plane (i.e. in perpendicular orientation). From this plot it is clear that, although the *c*-axis $I_c$ field dependence is steeper in the BZO sample, its $I_c$ values are higher in the whole field range up to 31 T. More importantly, the critical current at 5° off the *ab*-plane is significantly higher for the BZO sample.

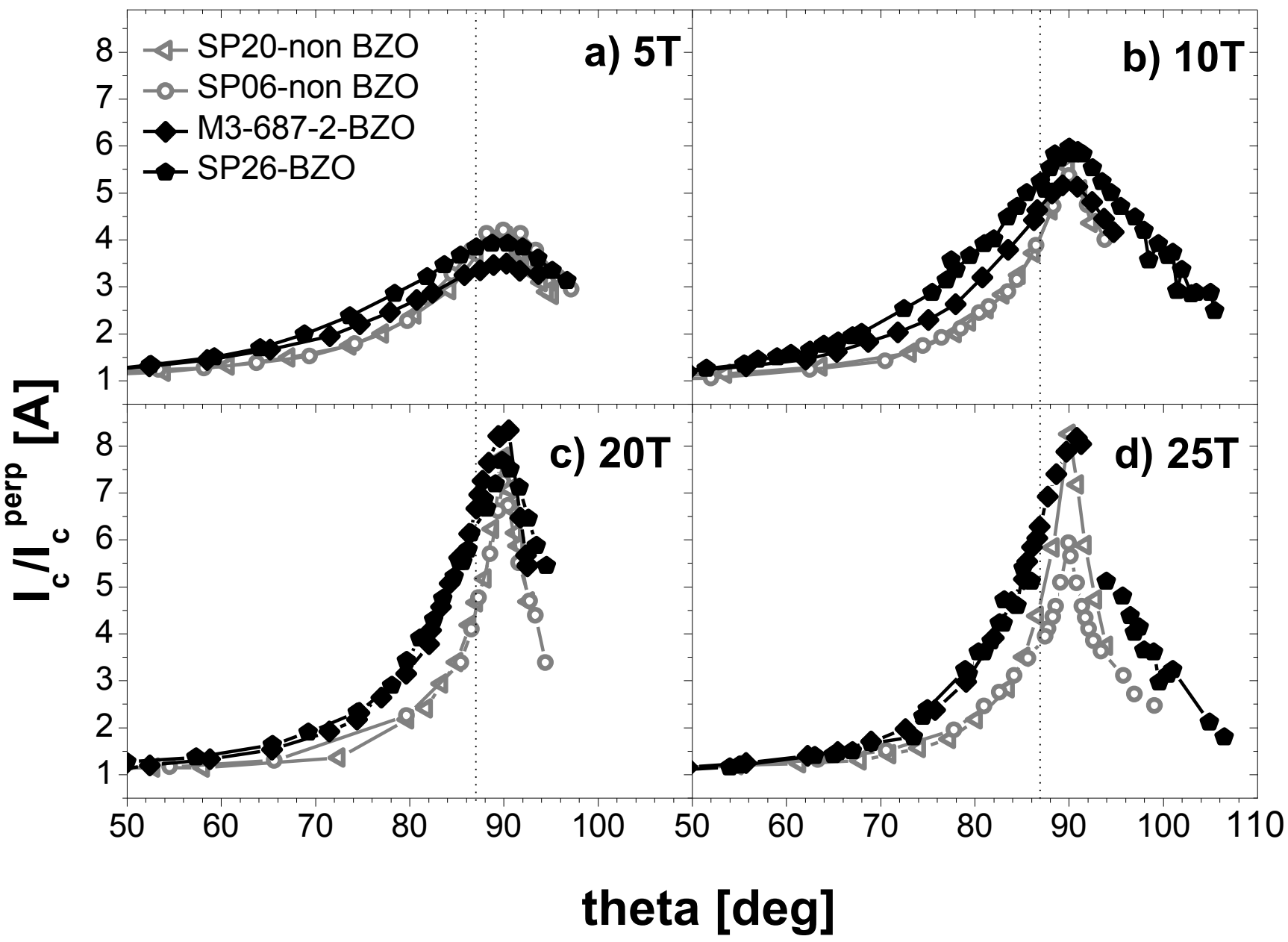

**Figure 6** $I_c$ normalized to $I_c$ perpendicular to the *ab* plane versus angle for the samples SP20, SP06, M3-687-2 and SP26 at (a) 5 T, (b) 10 T, (c) 20 T and (d) 25 T.

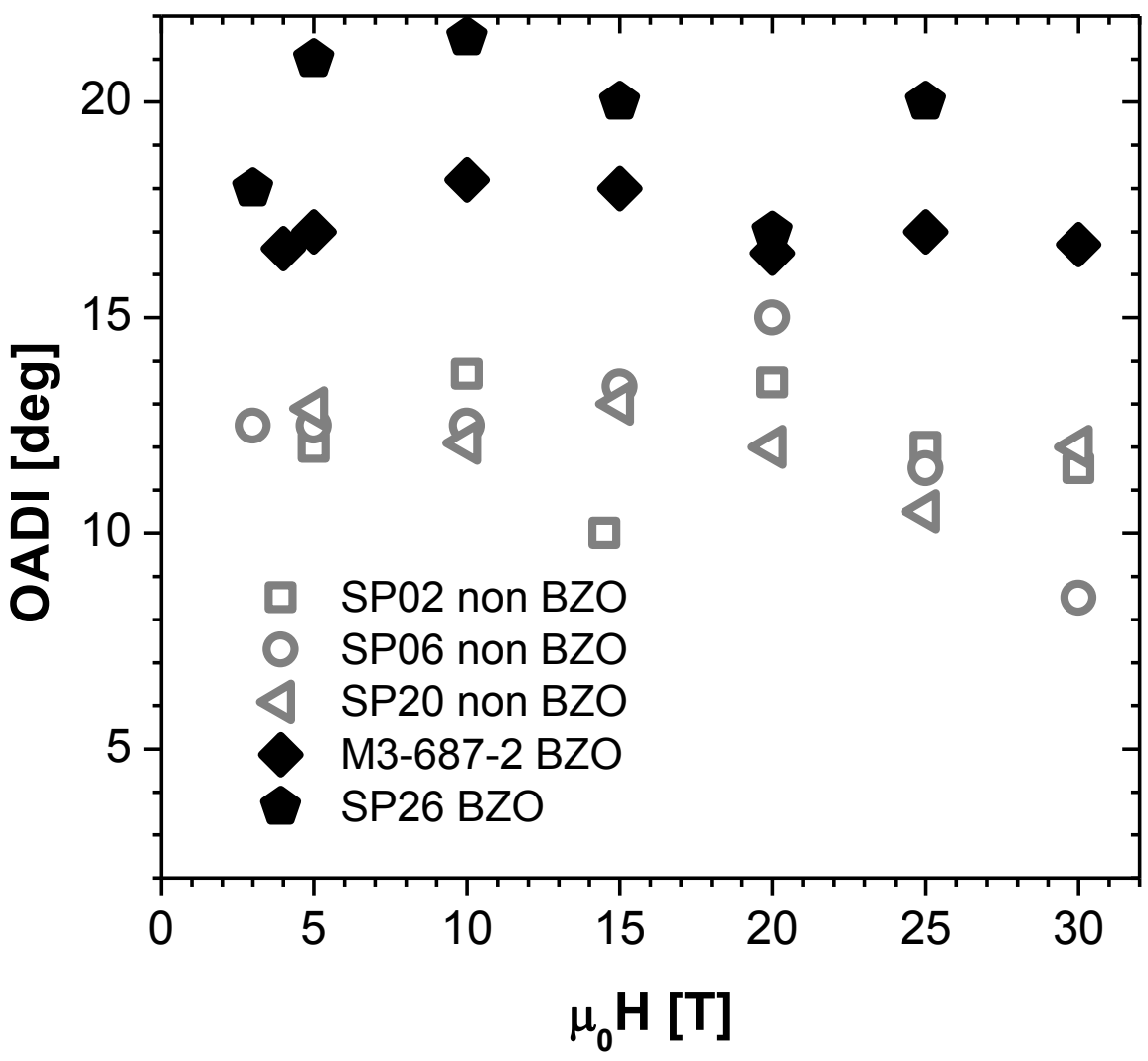

**Figure 7** Off-Axis Double $I_c$ (OADI) – i.e. the angle off the *ab* plane at which $I_c$ is twice the $I_c$ value perpendicular to the *ab* plane - vs. magnetic field.

Figure 9 shows the in-field dependence of $I_c$ for five samples with and without BZO at fixed angles from the *ab* plane. We can see that the latest generation of BZO CCs shows very good performance in terms of both absolute critical current values and angular dependence, a 4 mm conductor with 1 μm thick REBCO carrying about 700 A at 30 T when the field is 5 degrees off the *ab*-plane, an excellent result for magnet construction.

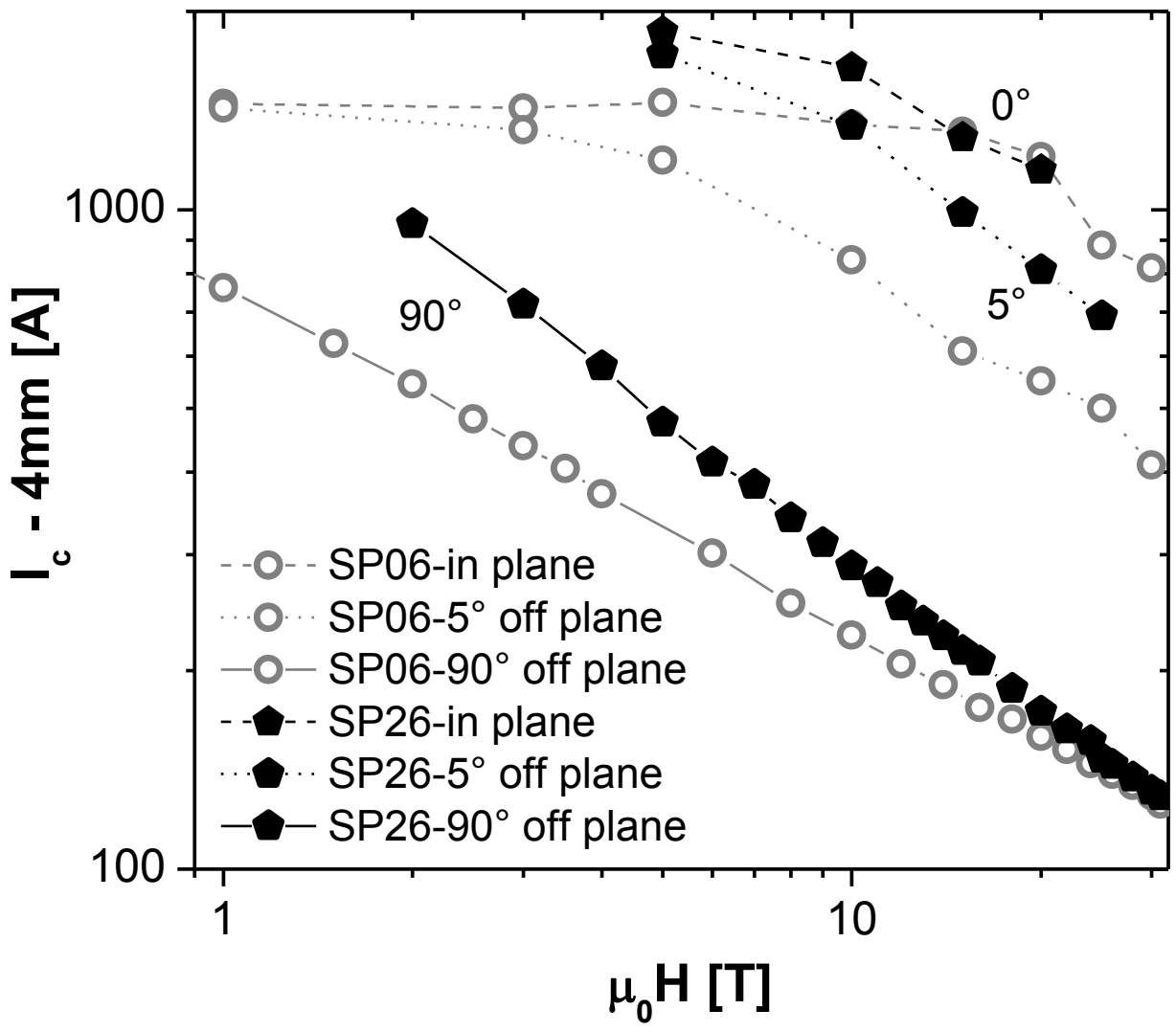

**Figure 8** Comparison of log $I_c$ vs. log $H$ in three different orientations for two samples without (SP06) and with (SP26) BZO nanorods. Angles are measured away from the tape plane.

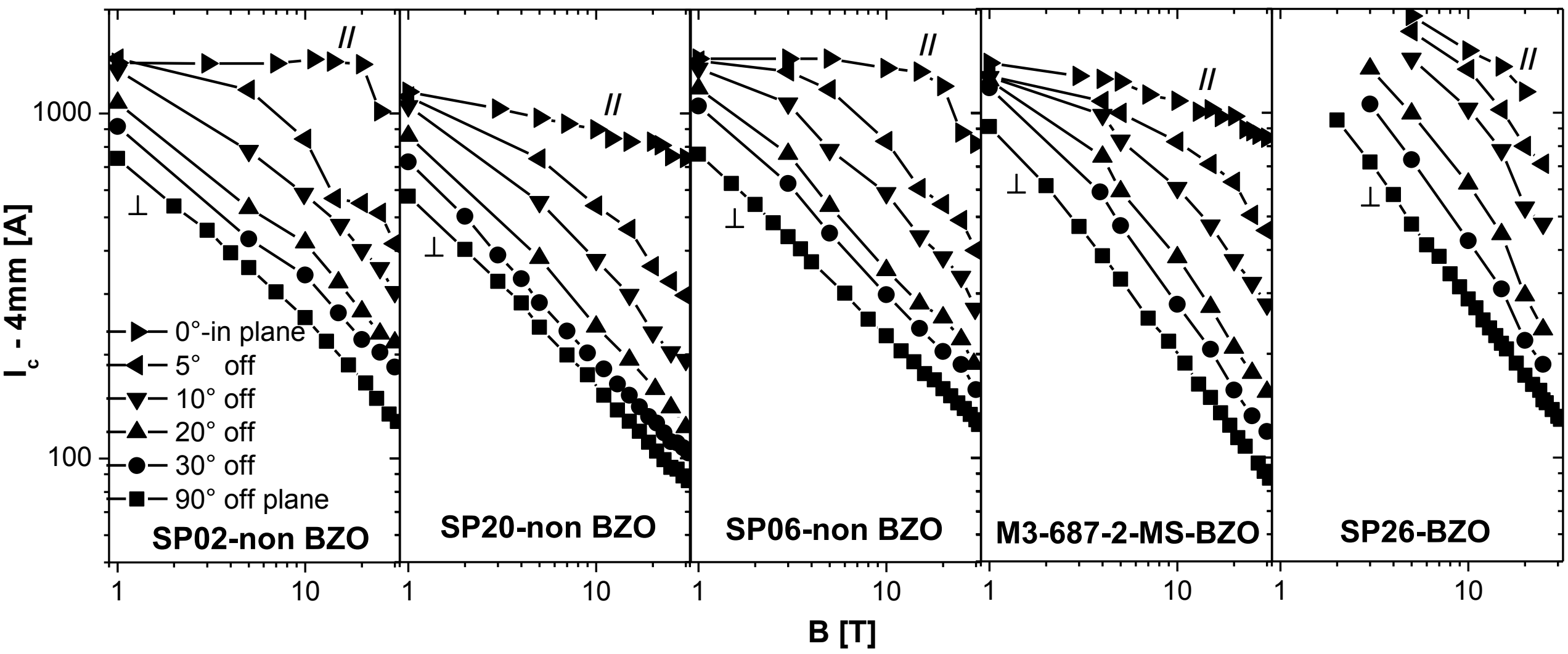

**Figure 9** $I_c$ vs. $B$ at fixed angles (here with respect to the *ab* plane) for 5 different CCs with or without BZO nanorods.

## 4. Discussion

The principal thrust of the work presented here has been to acquire data on the angular and field-dependent critical current behaviour of modern CCs over a broad field range at 4.2 K so as to assist the design of high field magnets. A simple and important conclusion of the work is that MOCVD conductors containing BZO nanorods are effective at all temperatures and fields up to at least 31 T at 4.2 K. This is a very positive conclusion that should avoid the need for production of different composition CCs for low and high temperature use. We are separately writing up a more detailed study of the temperature dependence of $J_c(\theta)$ and the operative pinning mechanisms [16] and so will restrict our discussion here to largely practical points.

Because access to high fields is not always easily available, scaling of the critical current density that will allow extrapolation from lower fields is useful. In figure 10(a) we plot $I_c$ vs. the perpendicular field for the same CCs as in figure 4 on a double logarithmic scale. It is clear that the perpendicular critical current follows the power-law $I_c \propto H^{-\alpha}$ quite well for all the samples [17]. $\alpha$ is about 0.5 for the samples without Zr additions in the range 4-31 T, while the field dependence is much steeper ($\alpha \sim 0.7$) and occurs over a wider field range (between about 2-31 T) for the samples with BZO precipitates. However, although BZO samples show a steeper $J_c(H)$, the most recently produced sample (SP26) has higher $I_c$ than all the other samples in the entire field range. In figure 10(b) we plot $I_c / H^{-0.5}$ for the non-BZO samples and as $I_c / H^{-0.7}$ for the BZO samples so as to evaluate the constancy of $\alpha$. Actually, $\alpha$ is not exactly constant for either sample set, but an assumption of constant $\alpha$ is probably sufficiently predictable to allow extrapolation. It appears to vary more significantly for the non-BZO samples, perhaps due to the lack of strong *c*-axis correlated pinning, while for the BZO samples, $\alpha$ is much more constant, except for the early R&D sample. We are separately exploring the pinning mechanism responsible for this huge increase of isotropic pinning in the BZO samples at low temperature [16]. We estimate that it is principally due to point defects induced by strain produced by the precipitate-matrix mismatch between $RE_2O_3$ in the non-BZO and the $RE_2O_3$ and BZO in the Zr-containing samples. Both the enhanced $J_c$ and its stronger field dependence would be consistent with the much larger strains induced by BZO as opposed to $RE_2O_3$ [4,18]. The effects described in figure 10 are specifically for *H* perpendicular to the plane, but in fact they are also much more broadly valid.

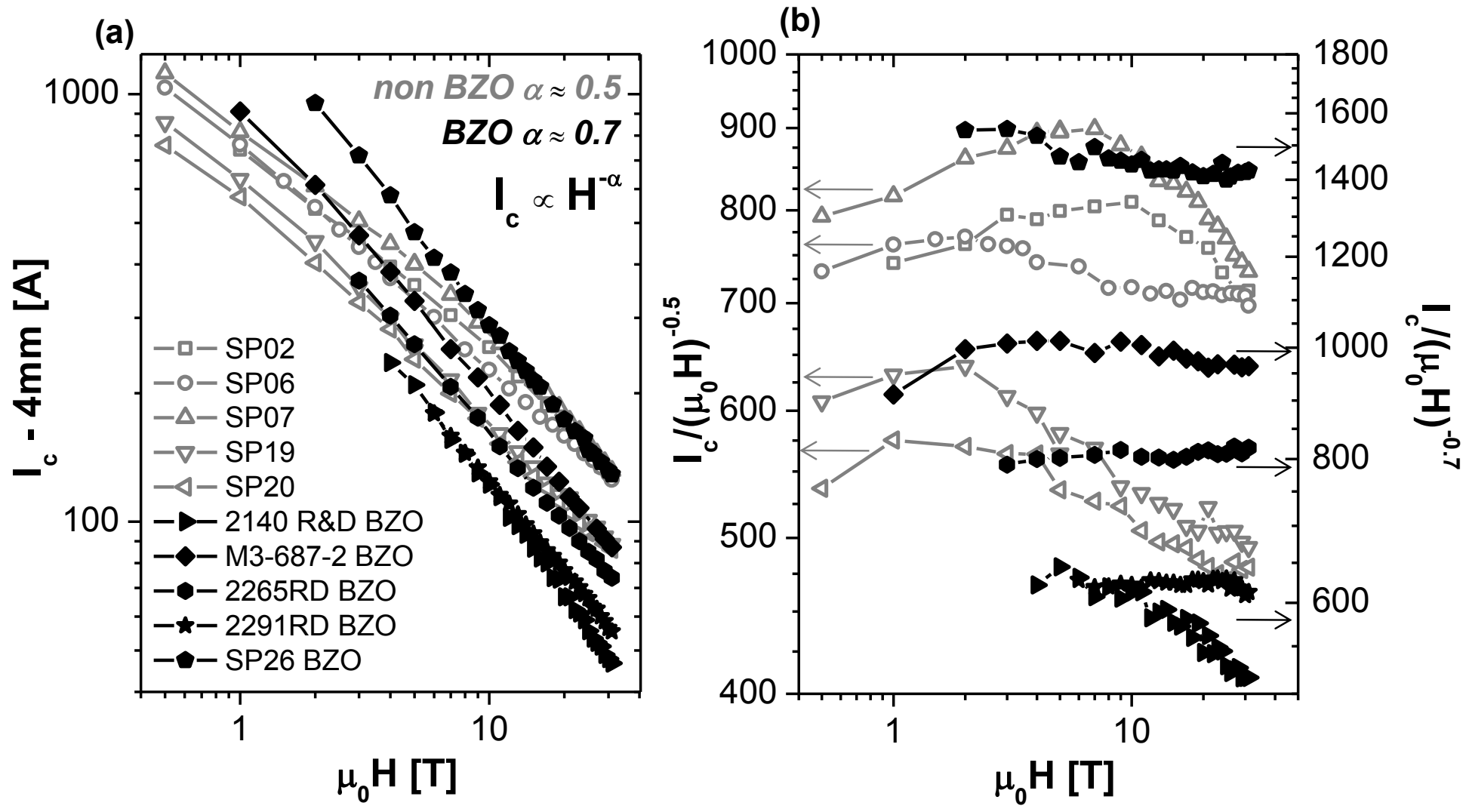

**Figure 10** (a) $I_c$ over 4 mm vs. magnetic field at 4.2 K in a log-log scale for several CCs without (gray hollow symbols) and with (black full symbols) BZO and (b) $I_c/(\mu_0 H)^{-0.5}$ (left axis) and $I_c/(\mu_0 H)^{-0.7}$ (right axis) vs. magnetic field for CCs respectively without and with BZO additions in configuration with *H* perpendicular to the CC plane.

At 77 K such a power-law field dependence has often been reported, using data taken immediately above the initial low-field plateau in which $J_c$ remains relatively constant. This power-law region usually occurs between 0.1 T and 1 T [2]. Typically [19], $\alpha$ values lie between 0.5 and 0.6 [20-23] for pure YBCO films. However, BZO reduces $\alpha$ to 0.3 [23] or 0.35-0.39 [24] and even 0.19 [19] because of its very strong *c*-axis correlated pinning effects. However, at low temperatures we observe quite opposite behaviour, $\alpha$ being ~0.5 for the non-BZO samples and ~0.7 for the BZO case. This is a quite specific indication that the influence of BZO on the pinning mechanisms at low and high temperature is completely different.

Regarding the angular dependence of $I_c$, it has been shown that at 77 K it dramatically depends on whether nanorods are present [2, 4, 19, 18]. Referring in particular to samples with Zr additions, BZO nanorods are effective strong pinning centres at high temperatures and low magnetic field because they strongly resist thermal fluctuations and are in the dense pinning limit where the density of pins, here about 2.5 T (figure 3), is comparable to or greater than the density of vortices. Typically, when the angle between the field and the nanorods exceeds the accommodation angle, the pinning drops sharply, thus explaining the high *c*-axis $J_c$ peak [3, 11, 24]. At 4.2 K this striking influence of nanorod doping on the angular dependence of $I_c$ is gone, as it was first reported in [11], and is clearly shown in figures 5 and 6. This must mean that additional pinning mechanisms have been introduced by the BZO at low temperatures which remove the strongly anisotropic high temperature pinning seen with BZO. Usually the angular scaling follows a generalized anisotropic scaling introduced by Blatter *et al.* [17] and Civale *et al.* [22] for a range as high as 60º away from the tape plane. Unfortunately it is precisely the next 30º where magnet designers are most concerned and we thus defer further discussion of properties near the tape plane to later work. Looking back to figure 9, however, we do see that $\alpha$ remains constant from 0-60°, that is up to about 30° off the *ab*-plane, emphasizing that good scaling can be obtained over a broad angular range away from the *ab*-plane.

## 5. Conclusions

We presented critical current measurements of several IBAD-MOCVD CCs produced by Super-Power Inc., both BZO-free and containing BZO nanorods. In particular, we made $I_c$ measurements at 4.2 K over the full angular range in fields up to 31 T, so as to be relevant for very high field magnet design.

We found that the field dependence of such conductors obeys a power-law behaviour $J_c \propto H^{-\alpha}$ with $\alpha \sim 0.5$ for BZO-free conductors and $\alpha \sim 0.7$ in BZO samples. Although the magnetic field dependence $J_c(H)$ for H perpendicular to the tape plane is steeper, the critical current values are higher over the entire magnetic field range for the most recent BZO-containing CC. Regarding the angular dependence of $I_c$, we observed that it looks quite similar in the samples with or without BZO: in contrast to what is generally observed at 77 K, BZO nanorods do not produce any *c*-axis peak in the $J_c(\theta)$ curve at 4.2 K at any field, while a cusp-like *ab* peak emerges with increasing magnetic field in both cases. When BZO precipitates are present, the pinning is strongly enhanced and the tendency of the *ab*-plane $J_c(\theta)$ peak to become cusp-like is moderated: the *ab* peak is actually significantly widened, leading to a higher current 5°-20° off the *ab* plane where it is important for magnet construction.

## Acknowledgements

A portion of this work was performed at the National High Magnetic Field Laboratory, which is supported by NSF Cooperative Agreement DMR-0654118, by the State of Florida, and additionally by the DOE through the Office of Electricity Delivery and Energy Reliability. The work at University of Houston and SuperPower was partially supported by the U.S. Department of Energy under a contract through U.T. Battelle. We thank A. Gurevich for fruitful discussion of pinning issues and U. Trociewitz, H. Weijers and other members of the HTS coil group for input on conductor-related issues.